\documentclass[reprint,pre]{revtex4-2}

\usepackage[utf8]{inputenc}
\usepackage{amsmath,amsfonts}
\usepackage{graphicx}
\usepackage{textcomp}
\usepackage{listings}
\usepackage{xcolor}

\newcommand\figwidth{8.cm}

\newcommand{\figlet}[1]{{\sffamily #1}}

\begin{document}
%%%%%%%%%%%%%%%%%%%%%%%%%%%%%%%%%%%%%%%%%%%%

\title{Thermodynamics of bouncing grains}

\date{\today}

\author{O.~Devauchelle}
\email[]{devauchelle@ipgp.fr}
\affiliation{Université de Paris, Institut de Physique du Globe de Paris, 1 rue Jussieu, CNRS, 75238 Paris, France}

\author{P.~Popović}
\affiliation{Faculty of Physics, University of Belgrade, POB 44, 11001 Belgrade, Serbia}

\author{P.~Szymczak}
\affiliation{Institute of Theoretical Physics, Faculty of Physics, University of Warsaw, Pasteura 5, 02-093 Warsaw, Poland}

\author{A.~Abramian}
\affiliation{Sorbonne Université, CNRS, Institut Jean Le Rond d'Alembert, F-75005 Paris, France}

\author{A.~Lazarus}
\affiliation{Sorbonne Université, CNRS, Institut Jean Le Rond d'Alembert, F-75005 Paris, France}
\affiliation{Massachusetts Institute of Technology, Department of Mathematics, Cambridge, MA 02139, USA}

\begin{abstract} % abstract
When a horizontal plate vibrates strongly enough, it causes small particles such as sand grains to continually bounce on it and, over time, to diffuse across its surface. This phenomenon is the cause of the well-known Chladni figure, which is drawn by a higher density of grains gathering along the nodal lines of a resonating elastic plate. Using a heterogeneous, non-resonating plate, we investigate experimentally this type of diffusion. We find that, for the most part, is it comparable to classical molecular diffusion. We can define a temperature for the bouncing grains, and the system then obeys the fluctuation-dissipation theorem. We also recover Maxwell-Boltzmann statistics at equilibrium, when temperature is uniform. However, when temperature varies across the vibrating plate, the microscopic details of the grains' dynamics affect their macroscopic behavior: Fick's law, for instance, no longer applies. Instead,
our experiments support a new transport relation that was recently proposed to represent diffusion in Chladni's experiment. Finally, we propose an expression for the heat flux associated to the non-equilibrium steady state predicted by this new relation, and test it against observations.
\end{abstract}

%%%%%%%%%%%%%%%%%%%%%%%%%%%%%%%%%%%%%%%%%%%%
\maketitle
%%%%%%%%%%%%%%%%%%%%%%%%%%%%%%%%%%%%%%%%%%%%

%%%%%%%%%%%%%%%%%%%%%%%%%%%%%%%%%%%%%%%%%%%%
\section{Introduction\label{sec:intro}}
%%%%%%%%%%%%%%%%%%%%%%%%%%%%%%%%%%%%%%%%%%%%

The formation of the Chladni figure---the gathering of sand grains along the nodal lines of a vibrating plate \cite{chladni1787entdeckungen}---has long been attributed to an average force, of ambiguous origin, that would deterministically drive the particles \cite{faraday1831xvii,kudrolli2008swarming,arango2016stochastic}. The tracking of individual grains over a vibrating membrane, however, suggests a simpler explanation: The grains are random walkers which gather where their diffusivity $D$ is weak \citep{abramian2025chladni}, as first suggested by \citet{grabec2017vibration}.

Perhaps surprisingly, this simple phenomenon cannot be represented by Fick's law, which states that the average flux of particles is
\begin{equation}
  \mathbf{j} = - D \nabla \rho
  \label{eq:Fick}
\end{equation}
where $\rho$ is the average density of particles. In equilibrium, indeed, the Fickian flux must vanish, and the concentration is then uniform, regardless of any variation in diffusivity. This paradox was first notes by \citet{buttiker1987transport} and \citet{landauer1988motion} who suggested that Fick's law be replaced with
\begin{equation}
  \mathbf{j} = - \nabla ( D \rho ) \, ,
  \label{eq:new_Fick}
\end{equation}
under some conditions, which  could be those of a rarefied gas \cite{van1988relative,volpe2016effective,raja2024diffusive}. Then, when the flux of particle vanishes, the density of particles must be inversely proportional to diffusivity:
\begin{equation}
\rho \propto \dfrac{1}{D} \, . 
\label{eq:one_over_D}
\end{equation}
This relation is observed experimentally in Chladni figures \citep{abramian2025chladni}. Sand grains bouncing on a vibrated surface, therefore, can be seen as a macroscopic analogue of the thought experiment introduced by \citet{buttiker1987transport} and \citet{landauer1988motion}---an analogue in which, conveniently, the motion of individual particles can be experimentally tracked.

Hereafter, we call ``statistical equilibrium'' the steady state associated to equation \eqref{eq:one_over_D}, in reference to the absence of any particle flux ($\mathbf{j}=0$), and in contrast with true thermodynamic equilibrium, for which the energy flux also needs to vanish. \citet{buttiker1987transport} went beyond this statistical equilibrium. He remarked that, when combined with a suitable potential force, a periodic temperature field could sustain a constant current of particles, thus driving a Brownian motor, now referred to as a ``Büttiker-Landauer (BL) ratchet''---a non-equilibrium steady state that is \emph{not} in statistical equilibrium ($\mathbf{j}\neq 0$). The analogy with the Chladni figure implied that a macroscopic version of the BL motor could be made with bouncing grains. In a companion paper, we present such an experimental realization, and show that it behaves essentially as predicted by \citet{buttiker1987transport}: A constant, average current of particles emerges from the random bouncing of sand grains \cite{companionPRL}.

Over the years, the BL motor has attracted the attention of theoreticians, not only because the development of nanoscale devices makes its physical realization seem ever more accessible, but also because its thermodynamics is intriguing \cite{berger2009optimal}. The main difficulty in assessing the motor's efficiency is to evaluate the leakage of heat along the temperature gradient. \citet{benjamin2008inertial} addressed this issue both theoretically and with numerical molecular dynamics, but their conclusions remains to be experimentally tested---which is obviously challenging with molecular systems.

Here we use a bouncing-grain experiment to investigate the statistics and thermodynamics associated to equation~\eqref{eq:new_Fick}. The setup is the same as in the companion paper \cite{companionPRL}, but hereafter we focus on the steady state associated to equation~\eqref{eq:new_Fick}, wherein no particle current nor external force complicates the picture. It is thus a simpler system than the BL motor, but it nonetheless features the heat flux that dominates the energy balance of the motor. On a more conceptual level, this heat flux also keeps the system away from equilibrium, even when the particle flux $\mathbf{j}$ vanishes, and thus directly relates to the departure from Boltzmann's statistics that the Chladni figure requires.

We first consider the trajectories of the bouncing grains, and check classical results from statistical physics, such as the Maxwell-Boltzmann distribution of velocities and the fluctuation-dissipation relation (section~\ref{sec:bouncing_grains}). We then justify the use of equation~\eqref{eq:new_Fick} to represent the diffusion of grains bouncing over a heterogeneous substrate (section~\ref{sec:diffdiff}), and finally measure the heat flux associated to the statistical equilibrium of bouncing grains by tracking their individual trajectories (section~\ref{sec:heat}).

%%%%%%%%%%%%%%%%%%%%%%%%%%%%%%%%%%%%%%%%%%%%
\section{Bouncing grains\label{sec:bouncing_grains}}
%%%%%%%%%%%%%%%%%%%%%%%%%%%%%%%%%%%%%%%%%%%%

%%%%%%%%%%%%%%%%%%%%%%%%%%%%%%%%%%%%%%%%%%%%
\subsection{Experimental setup\label{sec:set_up}}
%%%%%%%%%%%%%%%%%%%%%%%%%%%%%%%%%%%%%%%%%%%%

\begin{figure}[t]
\includegraphics[width=\figwidth]{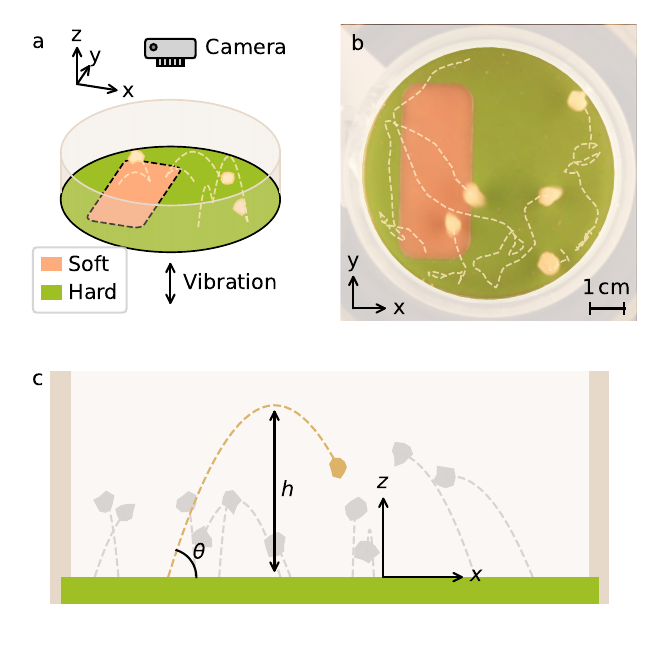}%
\caption{
\figlet{a}: Experimental setup. The disk vibrates along the $z$-axis. The setup can be tilted by a few degrees around the $x$-axis. Quartz grains bounce randomly over a hard plastic surface (green, PVC), or a soft silicon polymer (pink, vinyl polysiloxane).
\figlet{b}: Still frame from an experimental movie (run C). Dashed white lines show the grains' trajectories during $1.5\,$s (75 frames).
\figlet{c}: Idealized representation of the bouncing grains. Their trajectory during a jump is a free fall (dashed lines).
\label{fig:setup}}%
\end{figure}

\begin{figure}[t]
\includegraphics[width=\figwidth]{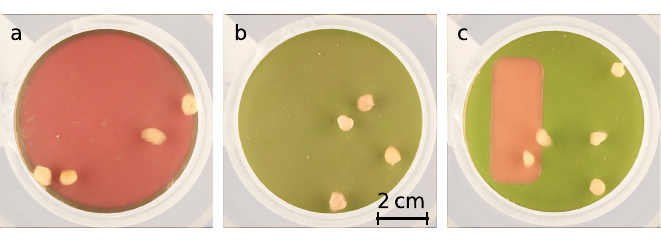}%
\caption{
The three substrates used in the experiment.
\figlet{a}: Soft elastomer (pink, VPS).
\figlet{b}: Hard plastic (green, PVC).
\figlet{c}: Heterogeneous substrate.
\label{fig:substrate}}%
\end{figure}

Inspired by the historical experiments of \citet{chladni1787entdeckungen}, and more recent versions of it \cite{grabec2017vibration,abramian2025chladni}, we devise a setup in which the vertical vibration of a plate causes irregular quartz grains to bounce erratically (figure~\ref{fig:setup}\figlet{a},\figlet{b}). The grains thus roam across a horizontal disk (radius $R=3.9\,$cm), made of rigid PVC and bounded by a 2$\,$cm-high PMMA wall which confines the grains over the disk. 
The surface of the disk is either simply painted green, and thus remains hard, or machine tooled with a miller to carve a 1$\,$mm-deep groove into which vinyl polysiloxane (VPS), a soft elastomer, is molded (pink area in figure~\ref{fig:substrate}). This grove can cover the entire surface of the disk (figure~\ref{fig:substrate}\figlet{a}), be absent (figure~\ref{fig:substrate}\figlet{b}), or cover only a small rectangle (heterogeneous substrate, figure~\ref{fig:substrate}\figlet{c}).
This allows us to alter the bounciness of the substrate, and thus cause the grains to behave differently over different areas of the vibrating disk. To keep the grains from sticking on the elastomer, we cover all disks with Parafilm M.

The PVC disk is fixed on a vertical aluminum shaft, the motion of which is constrained by two translating positioners that can only move along an optical rail. The verticality of this translation device, and therefore the horizontality of the vibrating disk, can be adjusted manually with a three-axes rotation stage. A disposable plastic cup connects the lower end of the shaft to the membrane of a bass loudspeaker (8$\,\Omega$), which is driven by a function generator (Hameg HM8030-6) through a 40$\,$W amplifier (Kemo M034N). The typical vibration frequency $\nu$ is between 30 and 40$\,$Hz.

\begin{table*}
\begin{tabular}{p{42mm}|cccccccc}
Run & A & B & C & D & E & F & G & H \\
 \hline 
Frequency \hfill  $\nu$ [Hz] & 30.6 & 30.4 & 29.8 & 29.8 & 40.6 & 40.6 & 30.4 & 29.8 \\
Grain size \hfill  $d_s$ [mm] & [5, 5.5] & [2.5, 4] & [4, 5] & [2, 2.5] & [4, 4.5] & [3, 4] & [2, 2.5] & [2, 2.5] \\
Tilt range \hfill $\phi$ [$^\circ$] & [-1.6,$\,$1.6] & [-0.9,$\,$1.3] & [-1.5,$\,$1.7] & [-1.0,$\,$1.1] & [-2.1,$\,$1.2] & [-1.5,$\,$1.8] & [-1.0,$\,$1.2] & [-0.9,$\,$1.2] \\
Number of movies & 60 & 30 & 60 & 30 & 20 & 30 & 30 & 30 \\
\hline
Diffusivity &  &  &  &  &  &  &  &  \\
 \hfill $D_{\mathrm{hard}}$  [cm$^2\,$s$^{-1}$] & 2.3$\,$\textcolor{gray}{\footnotesize $\pm$0.7} & 2.9$\,$\textcolor{gray}{\footnotesize $\pm$0.5} & 1.8$\,$\textcolor{gray}{\footnotesize $\pm$0.8} & 1.8$\,$\textcolor{gray}{\footnotesize $\pm$0.6} & 0.9$\,$\textcolor{gray}{\footnotesize $\pm$0.4} & 1.4$\,$\textcolor{gray}{\footnotesize $\pm$0.3} & 2.3$\,$\textcolor{gray}{\footnotesize $\pm$0.3} & 2.8$\,$\textcolor{gray}{\footnotesize $\pm$0.7} \\
 \hfill $D_{\mathrm{soft}}$  [cm$^2\,$s$^{-1}$] & 4.4$\,$\textcolor{gray}{\footnotesize $\pm$3.2} & 3.9$\,$\textcolor{gray}{\footnotesize $\pm$3.1} & 5.4$\,$\textcolor{gray}{\footnotesize $\pm$3.7} & 3.5$\,$\textcolor{gray}{\footnotesize $\pm$2.2} & 3.3$\,$\textcolor{gray}{\footnotesize $\pm$1.7} & 3.8$\,$\textcolor{gray}{\footnotesize $\pm$2.1} & 3.0$\,$\textcolor{gray}{\footnotesize $\pm$2.0} & 4.6$\,$\textcolor{gray}{\footnotesize $\pm$2.9} \\
\hline
Horizontal velocity (RMS) &  &  &  &  &  &  &  &  \\
 \hfill $v_{\mathrm{rms,\,hard}}$  [cm$\,$s$^{-1}$] & 10.5$\,$\textcolor{gray}{\footnotesize $\pm$3.1} & 10.7$\,$\textcolor{gray}{\footnotesize $\pm$2.0} & 9.0$\,$\textcolor{gray}{\footnotesize $\pm$4.6} & 9.2$\,$\textcolor{gray}{\footnotesize $\pm$3.3} & 6.8$\,$\textcolor{gray}{\footnotesize $\pm$2.4} & 8.4$\,$\textcolor{gray}{\footnotesize $\pm$3.6} & 9.9$\,$\textcolor{gray}{\footnotesize $\pm$1.6} & 11.3$\,$\textcolor{gray}{\footnotesize $\pm$3.5} \\
 \hfill $v_{\mathrm{rms,\,soft}}$  [cm$\,$s$^{-1}$] & 14.6$\,$\textcolor{gray}{\footnotesize $\pm$3.2} & 12.8$\,$\textcolor{gray}{\footnotesize $\pm$3.0} & 14.5$\,$\textcolor{gray}{\footnotesize $\pm$4.0} & 12.2$\,$\textcolor{gray}{\footnotesize $\pm$1.9} & 11.7$\,$\textcolor{gray}{\footnotesize $\pm$3.4} & 11.9$\,$\textcolor{gray}{\footnotesize $\pm$3.4} & 11.4$\,$\textcolor{gray}{\footnotesize $\pm$2.6} & 13.9$\,$\textcolor{gray}{\footnotesize $\pm$2.6} \\
 \hline 
\end{tabular}\\
\caption{Experimental parameters for the first series of experiments. The substrate is heterogeneous (figure~\ref{fig:substrate}\figlet{c}). The corresponding data is used in the companion paper \cite{companionPRL}. The vibration amplitude, which is constant during each run, was not recorded for this series. Brackets indicate sieve size for $d_s$. Numbers in gray indicate difference between autocorrelation and dispersion for $D$, and standard deviation, computed over a series of movies, for $v_{\mathrm{rms}}$.\label{tab:exp_params_first}}
\end{table*}

The particles we deposit on the vibrating plate are sand grains, mostly made of quartz, sieved into size classes (tables~\ref{tab:exp_params_first} and~\ref{tab:exp_params_acc}). For each experimental run, we introduce a small number of grains (five to ten) from a single class. The variability of their size is reported in tables~\ref{tab:exp_params_first} and~\ref{tab:exp_params_acc} as the mesh size of the two sieves in between which the grains of a given class are collected (except for runs I, L, M and N, for which the grains were hand picked, and their size estimated from pictures). The grains' roughness causes them to bounce in random directions---by contrast, smooth beads tend to maintain their direction over a few jumps. As they bounce over the vibrating plate, the grains typically jump up to a few millimeters above the plate's surface, but we did not measure this height directly.

We used a top-view camera (Canon EOS 600D) to record the horizontal trajectories of the grains with a series of one-minute long movies at a frame rate of 50$\,$Hz, which is just fast enough to keep a clear view of individual grains (figure~\ref{fig:setup}\figlet{b}). The camera is controlled with the Linux library \texttt{gPhoto2} \cite{gphoto}. It records images of size $1280 \times 720\,$px, which corresponds to a spatial resolution of $0.14\,$mm$\,$px$^{-1}$ on the vibrating surface. The grains appear as whitish over a green or pink background, and we can thus use the saturation of the image as an indicator of the grain's presence. After smoothing numerically the saturation frame over a distance of one grain size (Gaussian filter), the grains' centers appear as clear maxima. To track individual grains, we need to keep their number low enough that their identity from one image to the next remains unambiguous (about 5 grains for each experiments in practice). Then, we finally use the Munkres algorithm of the \texttt{scikit-image} Python library to reconstruct the trajectories of every grain for the duration of a movie (dashed lines in figure~\ref{fig:setup}\figlet{b}) \cite{munkres,scikit-image,browntrack}. Trajectories are broken when two grains come close to each other, but these events are rare enough to leave us with many long trajectories.

Finally, the loudspeaker, vibrating plate and camera are all fixed on a vertical rectangular frame held by two hinges which can rotate about the $x$ axis of figure~\ref{fig:setup}\figlet{a}. A stepper motor (Zaber NM23A200), controlled by an Arduino Motor Shield, turns a micrometric screw which pushes the frame away from the vertical, and thus tilts the vibrating disk with respect to the horizontal. We digitally record this tilt with a capacitive inclinometer (Meiri ME 26410), the output of which is converted numerically at a resolution of 0.01$^{\circ}$ (Microchip MCP3424). This inclination adds a tunable potential force to the grains' dynamics, which makes this setup suitable to run a BL motor \cite{companionPRL}. Here, we will use this feature only to measure the mobility of the bouncing grains.

During a typical experiment, we start the function generator, and observe the motion of four or five grains over the vibrating disk. We increase the vibration amplitude until the grains start to move randomly (when the amplitude is too low, the grains appear to hoover smoothly over the disk's surface \cite{kudrolli2008swarming}). We then check with the naked eye that the grains' density is symmetric about the rotation axis of the frame ($x$ axis), which allows us to finely tune the tilt of the disk and find its horizontal position. This manual procedure gave us better results than trying to find the zero of the inclinometer. We then record the number of movies we wish to use for these experimental conditions (typically a series of five movies, with a 30$\,$s pause between them). Finally, we change the vibration amplitude or tilt, and repeat.

In the present paper, we use two distinct series of experiments. During the first series (runs A to H, table~\ref{tab:exp_params_first}, also used in the companion paper~\cite{companionPRL}), we did not measure the vibration amplitude $A$, but it remained visibly less than 1$\,$mm. These runs featured a heterogeneous substrate, as shown on figure~\ref{fig:substrate}\figlet{c}, over which the statistical equilibrium we investigate in the present paper can establish itself. The setup was tilted to different angles, and the combination of a heterogeneous substrate and an external force formed a BL motor---the subject of the companion paper \cite{companionPRL}. Here, we use this external force only in section~\ref{sec:flucdiss}, to measure the mobility of the bouncing grains.

Runs I to N (table~\ref{tab:exp_params_second}), by contrast, were specifically devised for the investigation of the grains' microscopic dynamics. They were made on a homogeneous substrate (hard or soft, figure~\ref{fig:substrate}\figlet{a,b}), which allowed better diffusivity measurements, especially over the soft polymer which,  in runs A to H, extended only over a small area. The vibrating disk remained horizontal in runs I to N, and we also affixed a digital accelerometer on the vibrating disk (Joy-It MMA8452Q), and measured the vibration amplitude to be typically about $A\sim 0.5\,$mm (table~\ref{tab:exp_params_second}). The dimensionless acceleration of the plate, $A\omega^2/g$ where $\omega = 2 \pi \nu$ is the pulsation and $g$ is the acceleration of gravity, was thus of order one in all our experiments. This combination of amplitude and frequency was set by trial an error but, in retrospect, it can be justified as follows. If the dimensionless acceleration is less than one, the grains do not leave the plate, and  the diffusivity vanishes. Conversely, far above this threshold, the diffusivity depends only weakly on the type of substrate \cite{abramian2025chladni}, thus impairing the heterogeneity we are after.

In the next section, we introduce a simple model to represent the motion of the bouncing grains, and test it against these two data sets.

\begin{table*}
\begin{tabular}{p{40mm}|cccccc}
Run & I & J & K & L & M & N \\
 \hline 
Substrate & hard & soft & hard & soft & hard & soft \\
Grain size \hfill  $d_s$ [mm] & 1.5$\,$\textcolor{gray}{\footnotesize $\pm$0.2} & [2.5, 4] & [2.5, 4] & 6$\,$\textcolor{gray}{\footnotesize $\pm$0.6} & 6$\,$\textcolor{gray}{\footnotesize $\pm$0.6} & 1.5$\,$\textcolor{gray}{\footnotesize $\pm$0.2} \\
Number of movies & 41 & 24 & 55 & 42 & 41 & 50 \\
Acceleration range \hfill   $A\omega^2/g$ & [1.6, 2.7] & [1.4, 2.1] & [0.8, 2.7] & [1.3, 2.5] & [1.7, 2.9] & [1.5, 1.8] \\
 \hline 
\end{tabular}
\caption{Experimental parameters for the second series of experiments. The substrate is either soft or hard, but always uniform (figure~\ref{fig:substrate}\figlet{a,b}). The disk was kept horizontal for these runs, and vibrated at a frequency of 30$\,$Hz. The vibration amplitude was varied, diffusivity and root mean square velocity changed accordingly. Brackets indicate sieves size for $d_s$, whereas gray numbers indicate size variability for hand-picked grains. \label{tab:exp_params_acc}\label{tab:exp_params_second}}
\end{table*}

%%%%%%%%%%%%%%%%%%%%%%%%%%%%%%%%%%%%%%%%%%%%
\subsection{Ballistic trajectory \label{sec:free_fall} \& Random walk\label{sec:random_walk}}
%%%%%%%%%%%%%%%%%%%%%%%%%%%%%%%%%%%%%%%%%%%%

\begin{figure}[t]
\includegraphics[width=\figwidth]{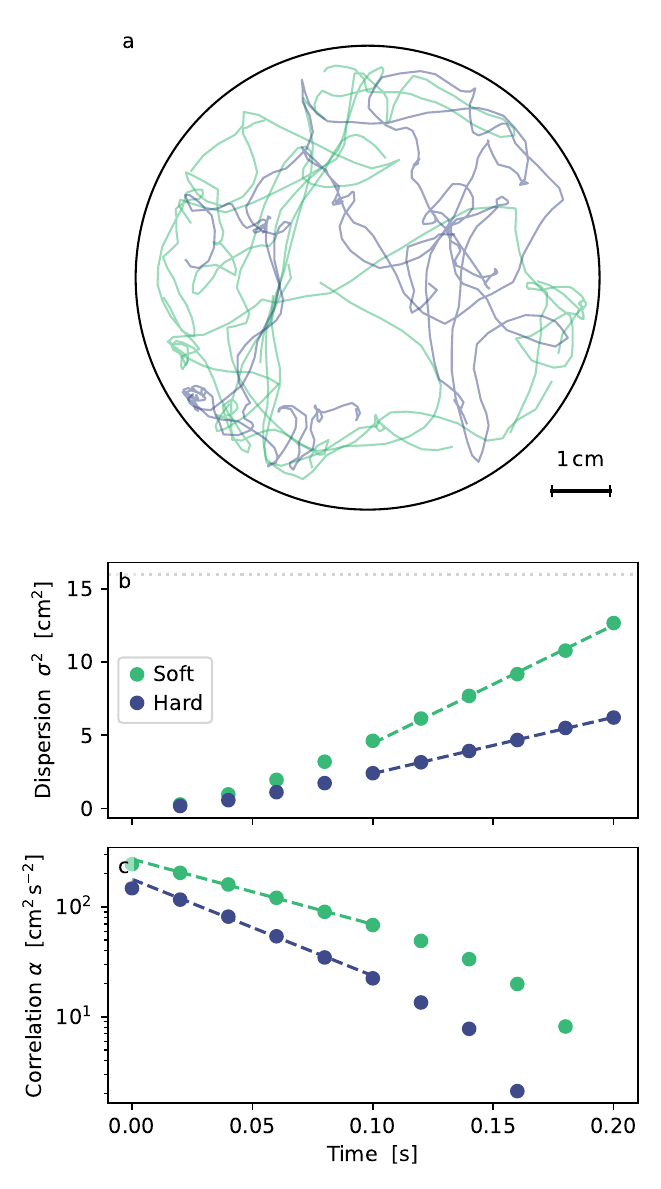}%
\caption{
\figlet{a}: Examples of trajectories over hard and soft disk (runs L and M, $f=30\,$Hz, $d_s=6\,$mm, dimensionless acceleration $A\omega^2/g=1.6$ and 1.9 respectively).
\figlet{b}: Average dispersion $\sigma^2$ of trajectories with time (dots). Dashed lines show linear fits with slope $4D$. Hard: $D=9.6\,$cm$^2\,$s$^{-1}$; Soft: $D=20.2\,$cm$^2\,$s$^{-1}$. Dotted gray line indicates box size $R^2$.
\figlet{c}: Autocorrelation function of the grains' velocity (dots). Dashed line shows exponential fit, with amplitude from equation~\eqref{eq:Green_Kubo}. Hard: $D=8.8\,$cm$^2\,$s$^{-1}$, $\tau_c=0.049\,$s; Soft: $D=19.9\,$cm$^2\,$s$^{-1}$, $\tau_c=0.073\,$s.
\label{fig:random_walk_autocorrelation}}%
\end{figure}

When the vertical acceleration of the vibrating plate exceeds that of gravity ($A\omega^2 \gtrsim g$), the grains start bouncing over its surface, in a fashion that is similar to what happens in a regular Chladni experiment \cite{grabec2017vibration,abramian2025chladni}. To the naked eye, the frequency of their bouncing is lower than that of the vibrating disk, and their horizontal motion seems erratic. The latter is confirmed by particle tracking (figure~\ref{fig:random_walk_autocorrelation}\figlet{a}): The horizontal trajectory of a grain is irregular and, over time, seems to explore the entire disk over which it is confined.

To ground this observation in measurements, we compute the mean square displacement of a collection of trajectories, namely
\begin{equation}
  \sigma^2 = \left< \Vert \mathbf{x} - \mathbf{x_0} \Vert^2 \right>
\end{equation}
where $\mathbf{x}$ is the position of a grain at a given time, $\mathbf{x}_0$ its position at the onset of its trajectory, and $\langle \cdot \rangle$ is the ensemble average, which we approximate with the average over our data set. To mitigate the influence of the walls, we cut each trajectory wherever it comes too close to a wall, and treat the resulting pieces as independent trajectories. The square displacement $\sigma^2$ is averaged over many trajectories, and we can plot it as a function of its duration $t$ (figure~\ref{fig:random_walk_autocorrelation}\figlet{b}). As expected, the square displacement grows with time, and this growth becomes linear after about 0.05$\,$s. Beyond about 0.2$\,$s, however, the displacement becomes comparable to the box size (dotted line in figure~\ref{fig:random_walk_autocorrelation}\figlet{b}), and our measurements become meaningless beyond this point.

We interpret the linear regime as the signature of a random walk, and therefore the slope of this relation as a measurement of the corresponding diffusivity. For run M, for instance, we find $D = 9.6\,$cm$^2\,$s$^{-1}$ over a hard substrate. For comparison, run L is made under similar circumstances, but over a soft elastomer disk; we then find $D = 20.2\,$cm$^2\,$s$^{-1}$---about twice the diffusivity over the hard substrate. Diffusivity also depends on grain size and plate acceleration, but we will not investigate these dependencies here (appendix~\ref{sec:acceleration}).

If the random-walk interpretation is correct, then the faster-than-linear growth of the square displacement that occurs before the linear regime corresponds to the ballistic regime, that is, to the free trajectory of a grain before it hits the substrate. Accordingly, we expect that the correlation time $\tau_c$ of a trajectory be comparable to the cross-over time between these two regimes.

To check this, we now compute the  autocorrelation function of the horizontal velocity, $\alpha(t)= \langle v(t) \, v(0)\rangle$, and plot it as a function of time (figure~\ref{fig:random_walk_autocorrelation}\figlet{c}). We find that it can be represented by an exponential decay, at least for a correlation time of less than about 0.1$\,$s. The characteristic time of this decay is $\tau_c = 0.049\,$s for run M ($0.073\,$s for run L), which compares with the cross-over time visible in figure~\ref{fig:random_walk_autocorrelation}\figlet{b}. For completeness, we now use the autocorrelation function to estimate again the diffusivity, by invoking the Green-Kubo relation:
\begin{equation}
  D = \int_0^{\infty} \alpha(t) \, \mathrm{d}t \, .
  \label{eq:Green_Kubo}
\end{equation}
For run M, for instance, we find $D=8.8\,$cm$^2\,$s$^{-1}$ with this method, to be compared with $D=9.6\,$cm$^2\,$s$^{-1}$ from the linear fit of the dispersion as a function of time (figure~\ref{fig:random_walk_autocorrelation}\figlet{b},\figlet{c}). We generally obtain values less than 10$\,$\% away from the dispersion measurements, which indicates that, over a flat uniform surface, bouncing grains diffuse randomly. Based on the decorrelation time of their random walk, we expect that their jumps typically last a few hundredths of a second.

We now wish to interpret these observations within a simple theoretical framework, in which we will favor clarity over exactness. In particular, we assume that the randomness of the grain's motion is limited to their rebounds, during which the roughness of their surface diverts them from their initial horizontal course. In between jumps, accordingly, we expect their trajectory to follow a ballistic parabola. The duration $\tau$ of a jump of height $h$ then reads
\begin{equation}
  \tau =  2 \sqrt{\dfrac{2 h}{g}} \, .
  \label{eq:tau}
\end{equation}
Interpreting the autocorrelation decay time $\tau_c$ as the typical jump duration $\tau$, this expression yields a bouncing height of $h\sim3\,$mm. This order of magnitude matches the estimate that one can make with a naked eye.

Based on the same assumptions, the horizontal extension $\ell$ of a step is
\begin{equation}
  \ell = v \, \tau \, , 
  \label{eq:ell}
\end{equation}
where $v$ is the norm of the grain's horizontal velocity. In our simplistic model, this velocity is a constant, but of course in practice it varies from one bounce to the next. To get a sense of scales, we will later identify it with the root mean square velocity we measure in experiments, but this will only be a rough approximation. Returning to the model, we neglect the friction of air, and accordingly assume that energy is conserved as long as the grain is detached from the plate. We can thus relate the vertical velocity of the grain at the onset of a jump to the height $h$ of that jump:
\begin{equation}
  v = \dfrac{\sqrt{2 g h}}{\tan \theta}
    \label{eq:v}
\end{equation}
where $\theta$ is the initial angle of the trajectory with respect to a horizontal (figure~\ref{fig:setup}\figlet{c}). We do not expect any specific value for the angle $\theta$ based on general principles; we rather expect it to depend on microscopic details such as the shape of the grains. Despite its  mundane origin, this angle decides of the partition of energy between vertical and horizontal directions---equipartition requires $\theta=45^{\circ}$.

Overall, equations~\eqref{eq:tau} to \eqref{eq:v} leave us with two free parameters: The bouncing angle $\theta$, which we cannot measure directly, and the horizontal velocity of a grain $v$, which we can measure with particle tracking. We do so in the next section.

\subsection{Maxwell-Boltzmann distribution\label{sec:measurements}}

\begin{figure}[t]
\includegraphics[width=\figwidth]{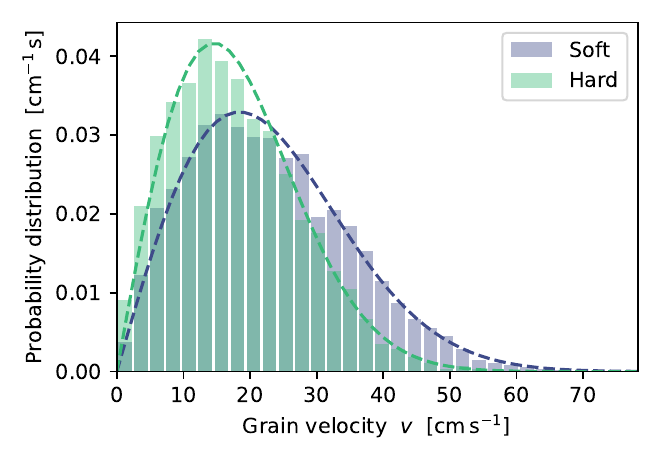}%
\caption{Velocity distribution over a uniform substrate (runs L and M, $f=30\,$Hz, $d_s=6\,$mm, dimensionless acceleration $A\omega^2/g=1.6$ and 1.9 respectively). Colors correspond to type of substrate. Bars: experimental probability distribution. Dashed lines: Maxwell-Boltzmann distribution with the same root mean square value ($v_{\mathrm{rms}}=26.1\,$cm$\,$s$^{-1}$ over the soft substrate, and $20.6\,$cm$\,$s$^{-1}$ over the hard substrate).
\label{fig:velocity}}%
\end{figure}

Using the trajectories resulting from the grain tracking procedure, we can compute the horizontal velocity $v$ of a grain between two camera frames. The frame rate of the camera is accurate, so the uncertainty of this measurement stems from the grains' position only. The latter is of the order of a pixel, which yields an uncertainty of about 1$\,$cm$\,$s$^{-1}$. Figure~\ref{fig:velocity} shows the distribution of the grains' velocity for experimental runs L and M. We find that these distributions are broad and, within the uncertainty of our measurements, they are well represented by a two-dimensional Maxwell-Boltzmann distribution, that is:
\begin{equation}
  \rho_v = \dfrac{2 v}{v^2_{\mathrm{rms}}} e^{ -\left( {v}/{v_{\mathrm{rms}}} \right)^2 } \, ,
\end{equation}
where $v_{\mathrm{rms}}=\sqrt{\langle v^2\rangle}$ is the root mean square of the horizontal velocity, measured for all the movies in an experimental run that were shot with the same vibrational acceleration. In other words, like in an ideal gas, the probability density of a grain in the two-dimensional velocity space decreases exponentially with its kinetic energy. Under otherwise similar conditions, the grains are faster over the soft substrate ($v_{\mathrm{rms}}\approx\,26.1\,$cm$\,$s$^{-1}$) than over the hard disk (20.6$\,$cm$\,$s$^{-1}$). In that sense, they are more energetic over the elastomer, as one would intuitively infer from their higher diffusivity (section~\ref{sec:random_walk}).

Invoking the energy balance during a jump, in the form of equation~\eqref{eq:v}, and identifying the horizontal velocity of our simple model with the root mean square velocity, we find that the bouncing angle is about $\theta\sim 50^{\circ}$ on the hard substrate (\citet{abramian2025chladni} measured $53^{\circ}$ in their Chladni experiment). We will get another estimate of this angle in section~\ref{sec:flucdiss}, but this value is certainly compatible with a the direct observation of our bouncing grains.

%%%%%%%%%%%%%%%%%%%%%%%%%%%%%%%%%%%%%%%%%%%%
\subsection{Temperature}\label{sec:flucdiss}
%%%%%%%%%%%%%%%%%%%%%%%%%%%%%%%%%%%%%%%%%%%%

\begin{figure}[t]
\includegraphics[width=\figwidth]{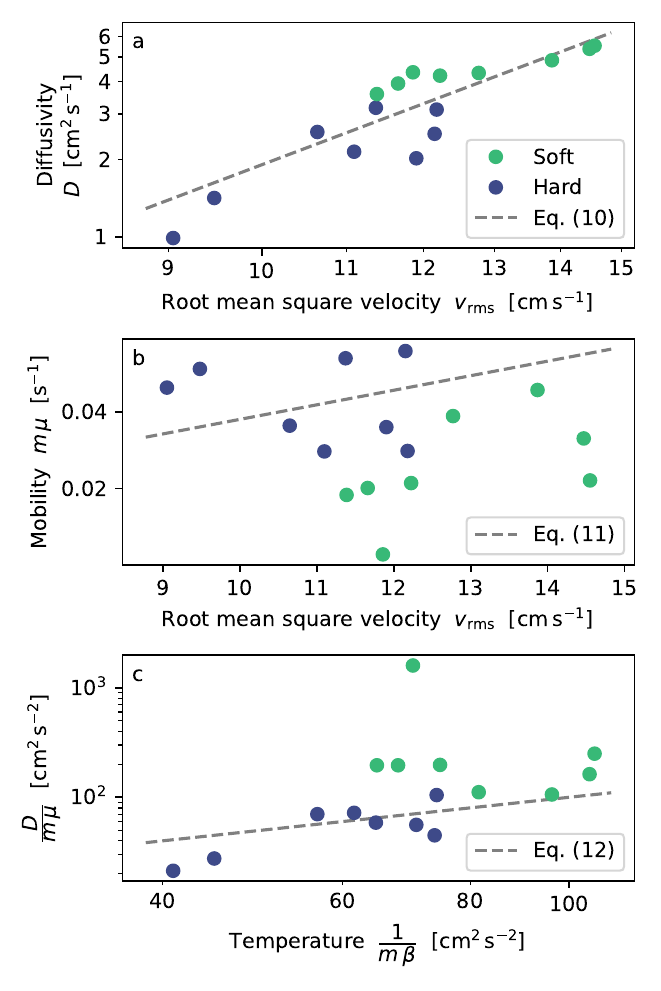}%
\caption{Empirical evaluation of the bouncing-grain model (section~\ref{sec:bouncing_grains}, runs A to H only). All scales are logarithmic.
\figlet{a}: Diffusivity as a function of root mean square velocity. Dashed line: equation~\eqref{eq:D_v} with a bouncing angle of $\theta = 75^{\circ}$ (fitted to data).
\figlet{b}: Mobility as a function of root mean square velocity. The product $m \mu$ is independent from the mass $m$ of a grain. Dashed line: equation~\eqref{eq:mu_v} with $\theta = 75^{\circ}$.
\figlet{c}: Fluctuation-dissipation relation. Dashed line: one-to-one relation, equation~\eqref{eq:fluc_diss}.
\label{fig:fluc_diss}}%
\end{figure}

We now turn to the randomness associated with a grain's bouncing. We hypothesize that a grain loses memory at each bounce, and that it takes off from the vibrating plate in any direction with equal probability. Projected on the vibrating plate, that is, on the ($x$,$y$) plane, the trajectory of a grain is thus a two-dimensional random walk with a correlation time $\tau$, and a characteristic step size $\ell=v \tau$. On average, this random walker diffuses over the plate with diffusivity
\begin{equation}
  D = \dfrac{\ell^2}{4 \tau}
    = \dfrac{v^3 \tan \theta }{2g} \, .
    \label{eq:D_v}
\end{equation}
We can test this expression with experimental grain trajectories, assuming that all its parameters are constant, including the bouncing angle $\theta$.

To do so, we use experimental runs A to H, wherein the grains bounce over a heterogeneous substrate. This allows us to measure the diffusivity $D$ and the root mean square velocity $v_{\mathrm{rms}}$ over the two substrates at exactly the same vibrational acceleration, provided we distribute the data into two bins corresponding to the hard and soft substrate. In addition, we changed the tilt angle of the substrate for these runs, and this will allow us to measure the mobility in the same runs.

Identifying the constant velocity of the model, $v$, with the root mean square velocity that we measure in experiments, $v_{\mathrm{rms}}$, we find that diffusivity consistently increases with velocity in runs A to H  (figure~\ref{fig:fluc_diss}\figlet{a}). The associated relation is compatible with an exponent of $3$ (a least mean square fit in log-log space yields an exponent of $3.3\pm 0.5$). Assuming the exponent is indeed $3$, we can fit equation~\eqref{eq:D_v} to the data, and we thus find a bouncing angle of $\theta \approx 75^{\circ}\pm 6^{\circ}$. This value is significantly different from the estimate of section~\ref{sec:measurements}.

Here we face the limitations of our order-of-magnitude model. The bouncing angle $\theta$ and the velocity $v$ vary from one bounce to the next, and should be thought of as random variables. Their mean or root mean square values, therefore, cannot be directly injected into equations~\eqref{eq:v} and \eqref{eq:D_v}---hence the discrepancy between the two estimates. A better model wherein each parameter has its own distribution can be envisioned, but our experimental setup gives us no access to the angle $\theta$, and such a model would therefore be overkill.
About the bouncing angle, we can only say with some confidence that its tangent is of order one. Unfortunately, this leaves us without any definitive answer about the equipartition of energy between the horizontal and vertical directions, which we cannot rule out. We would find it surprising, however, that equipartition be always satisfied, thus fixing the bouncing angle to 45$^{\circ}$ regardless of the grains' shape and the type of substrate.

\begin{figure}[t]
\includegraphics[width=\figwidth]{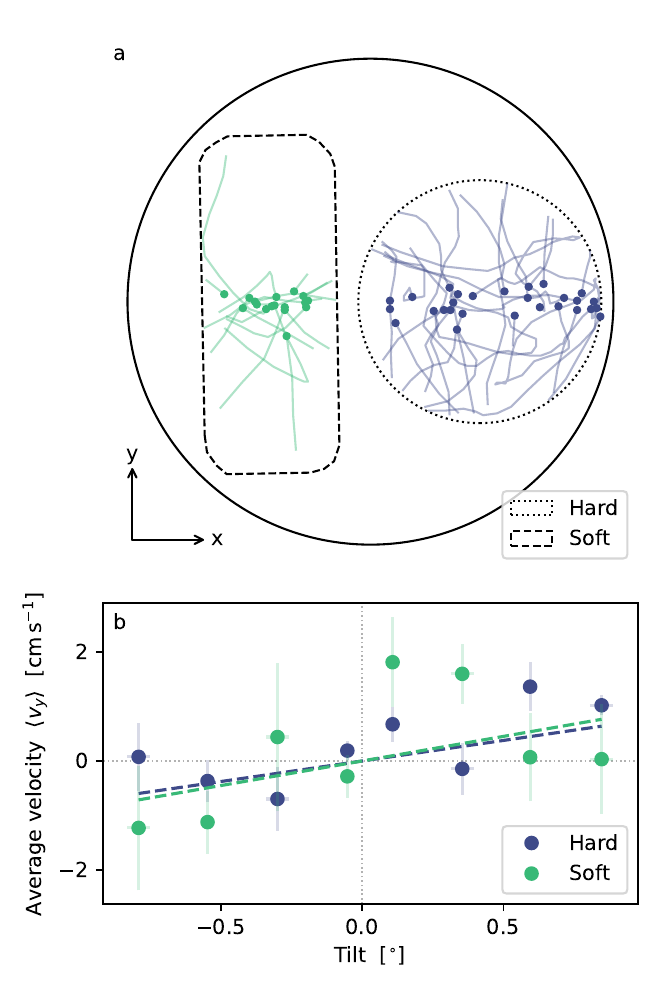}%
\caption{Measurement of mobility in the experiment.
\figlet{a}: Examples of truncated trajectories, all starting near the $y=0$ line, and split into two subdomains corresponding to hard and soft substrates (run C). Each trajectory starts from the dot.
\figlet{b}: Average velocity of the truncated trajectories as a function of the tilt angle for run C. For each domain (soft and hard) the data are distributed into nine bins and averaged therein. Error bars show standard deviation. The slope of the linear fit to the data is an estimate of the mobility $\mu$ (dashed lines). We find $mg \mu =52\,$cm$\,$s$^{-1}$ over the soft pad, and $mg \mu =43\,$cm$\,$s$^{-1}$ over the hard disk. 
\label{fig:mobility}}%
\end{figure}

To further the comparison with microscopic thermodynamics, we now investigate the inseparable companion  of diffusivity---mobility. If a horizontal force $f$ acts upon a grain of mass $m$, the latter will accelerate in response to the force as long as it is detached from the plate. Over the duration $\tau$ of its flight, it acquires an additional velocity equal to $f \tau/m$. Since we assume that, at each bounce, the grain's velocity is reset to zero, the acquired velocity averages over the duration $\tau$ of a flight to $f \tau/(2m)$. Defining the mobility $\mu$ of the grain as the ratio of this velocity to the force that induces it, we find
\begin{equation}
  m \mu = \dfrac{\tau}{2}
      = \dfrac{v}{g}  \tan \theta \, .
  \label{eq:mu_v}
\end{equation}
We can measure the mobility by applying a force on bouncing grains, and then track their trajectories. The ratio of the average velocity of a grain to the intensity of the force that drives it is the mobility.

To generate a horizontal force, we tilt the vibrating substrate with respect to a horizontal, and measure its angle (runs A to H). As a rough estimate of the average velocity $\langle v_y \rangle$ along the $y$ axis (the direction of the force), we first distribute the trajectories into two subdomains, namely the soft pad, and a circular domain of similar area over the hard disk (figure~\ref{fig:mobility}\figlet{a}). We now need to perform a Lagrangian measurement of the velocity (as opposed to the average velocity of a population of grains considered in a fixed domain). This measurement is valid only away from any boundary.

To get a Lagrangian average, we truncate the trajectories where they cross the $x$-axis, and then where they leave the domain of interest. Finally, we average the velocity over those truncated trajectories, and plot the resulting value as a function of the tilt of the plate (figure~\ref{fig:mobility}\figlet{b}). The data appear scattered around a linear trend, the slope of which is an estimate of mobility (actually, an estimate of $m g \mu$).

We are now ready to test equation~\eqref{eq:mu_v} against experiments provided, again, that we identify the typical velocity of a grain $v$ with the root mean square velocity $v_{\mathrm{rms}}$. We then plot the product $m \mu$ as a function of the root mean square velocity (figure~\ref{fig:fluc_diss}\figlet{b}). No free parameter remains at this stage, and we can directly plot the linear relation of equation~\eqref{eq:mu_v} on the same plot. For a bouncing angle $\theta = 75^{\circ}$, this expression appears compatible with the data, despite a large scatter. For unknown reasons, run F (soft substrate) appears as an outlier.

Finally, following classical convention, we define the macroscopic temperature $kT$ as the ratio of  diffusivity to  mobility:
\begin{equation}
  k T
    = \dfrac{D}{\mu}
    = \dfrac{m v^2}{2} \, .
  \label{eq:fluc_diss}
\end{equation}
Unsurprisingly, in the present theory, the macroscopic temperature of our bouncing grains is just their kinetic energy in the horizontal plane. This was to be expected for such a simple model; we now test this relation against observations.

Using the grains' trajectories, we estimate independently their diffusivity $D$, their mobility $\mu$ and their mean square velocity $v^2_{\mathrm{rms}}$. Identifying the latter as the temperature (up to an $m/2$ coefficient), we can test the fluctuation-dissipation relation, in the form of equation~\eqref{eq:fluc_diss}, without any reference to the bouncing angle (figure~\ref{fig:fluc_diss}\figlet{c}), nor to any other fitting parameter. Again, we find a large scatter around the line corresponding to the exact relation, with run~F still an outlier---the result, perhaps, of our fragile measurement of the mobility.

Overall, however, the data appear to accord with the fluctuation-dissipation relation. This allows us to consistently define a macroscopic temperature, either as the ratio of diffusivity to mobility, or as the average kinetic energy of the grains (in the plane of the vibrating substrate). In that sense, the bouncing grains are a sound analogue of molecular thermodynamics; they might prove useful, for instance, for the teaching of elementary statistical physics. Their macroscopic size, and the simplicity of the setup, might foster a developing intuition for thermodynamics.

The analogy with molecular thermodynamics, however, cannot be pushed too far. As we will see in section~\ref{sec:diffdiff}, indeed, it breaks down when the temperature is not uniform---and the diffusion law is at the core of this breakdown.

%%%%%%%%%%%%%%%%%%%%%%%%%%%%%%%%%%%%%%%%%%%%
\section{Itō diffusion\label{sec:diffdiff}}
%%%%%%%%%%%%%%%%%%%%%%%%%%%%%%%%%%%%%%%%%%%%

\subsection{Non-Fickian diffusion\label{sec:non-Fickian}}

\begin{figure}[t]
\includegraphics[width=\figwidth]{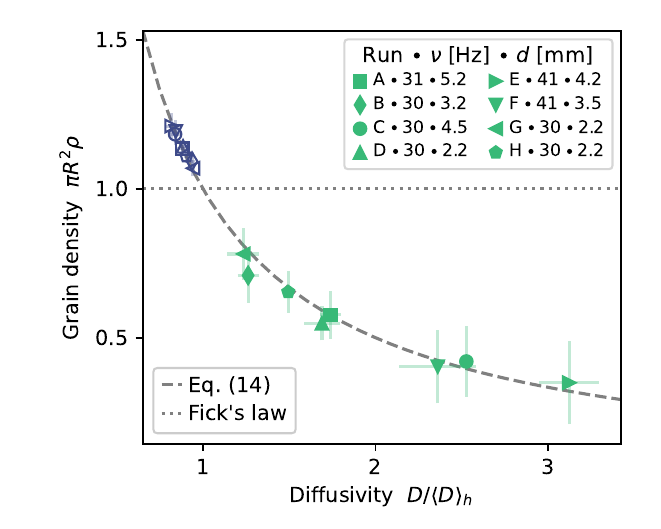}%
\caption{Grain density over a horizontal, heterogeneous substrate, as a function of dimensionless diffusivity. Each experimental run is represented by two data points (Green filled markers: soft pad; Empty blue markers: hard disk). Vertical error bars indicate standard deviation over successive movies. Horizontal error bar correspond to different estimates of the diffusivity ($2D = \langle y^2/t\rangle $ or $\langle y^2 \rangle/\langle t\rangle $).
\label{fig:one_over_D}}%
\end{figure}

So far, we only considered a random walk over a uniform substrate. Although we did use data from heterogeneous experiments, we split it into two substrate classes and treated these classes independently. We now investigate in details the behavior of bouncing grains over a heterogeneous substrate (figure~\ref{fig:substrate}\figlet{c}). Their diffusivity is now a function of space, and so is the temperature, since the two quantities are related by equations~\eqref{eq:D_v} and \eqref{eq:fluc_diss}, which we can recombine into
\begin{equation}
  kT = \dfrac{m}{2^{1/3}} \left( \dfrac{g D}{\tan \theta}\right)^{2/3} \, .
  \label{eq:kT_D}
\end{equation}
We thus exit the realm of thermodynamic equilibrium.

In the present paper (and its companion \cite{companionPRL}), we follow \citet{abramian2025chladni} and claim that equation~\eqref{eq:new_Fick} is a better representation of the diffusion of bouncing grains than Fick's law~\eqref{eq:Fick}. Although a formal derivation of this expression from first principles (i.e. from Newton's laws) is outside the scope of the present paper, we support this claim with three arguments. Namely,
(i) the inevitability of equation~\eqref{eq:new_Fick} for an explicit, unbiased random walk \cite{risken1989fokker,volpe2016effective};
(ii) the notion of bouncing, which implies that the external noise acts intermittently, at the onset of each jump (section~\ref{sec:bouncing_grains});
(iii) the heterogeneity of the statistical equilibrium reached in experiments, to which we now turn.

If Fick's law~\eqref{eq:Fick} were to hold on a heterogeneous substrate, the grains would distribute themselves uniformly in statistical equilibrium ($\mathbf{j}=0$). Instead, equation~\eqref{eq:new_Fick} yields
\begin{equation}
  \rho
    =  \dfrac{\langle \rho \rangle_a \langle D \rangle_h}{D}
    = \dfrac{\langle D \rangle_h}{\pi R^2 D}
  \label{eq:one_over_D_normed}
\end{equation}
at statistical equilibrium, where $\langle \cdot \rangle_h$ is the harmonic average over the vibrating disk, whereas  $\langle \cdot \rangle_a$ is the arithmetic average. Both are space averages, and $\langle \rho \rangle_a= 1/A$, where $A$ is the area of the domain. Telling the two theories apart is now straightforward---one simply needs to measure the density of the bouncing grains in an experiment at statistical equilibrium, and plot it as a function of diffusivity. Figure~\ref{fig:one_over_D} leaves no ambiguity as to which theory better accounts for observations (equation \eqref{eq:one_over_D_normed} requires no fitting parameter, and Fick's law predicts $\rho \pi R^2=1$).

In the next section, we investigate the difference between equation~\eqref{eq:new_Fick} and Fick's law, and the microscopic origin of these relationships.

\subsection{Drift\label{sec:drift}}

\begin{figure}[t]
\includegraphics[width=\figwidth]{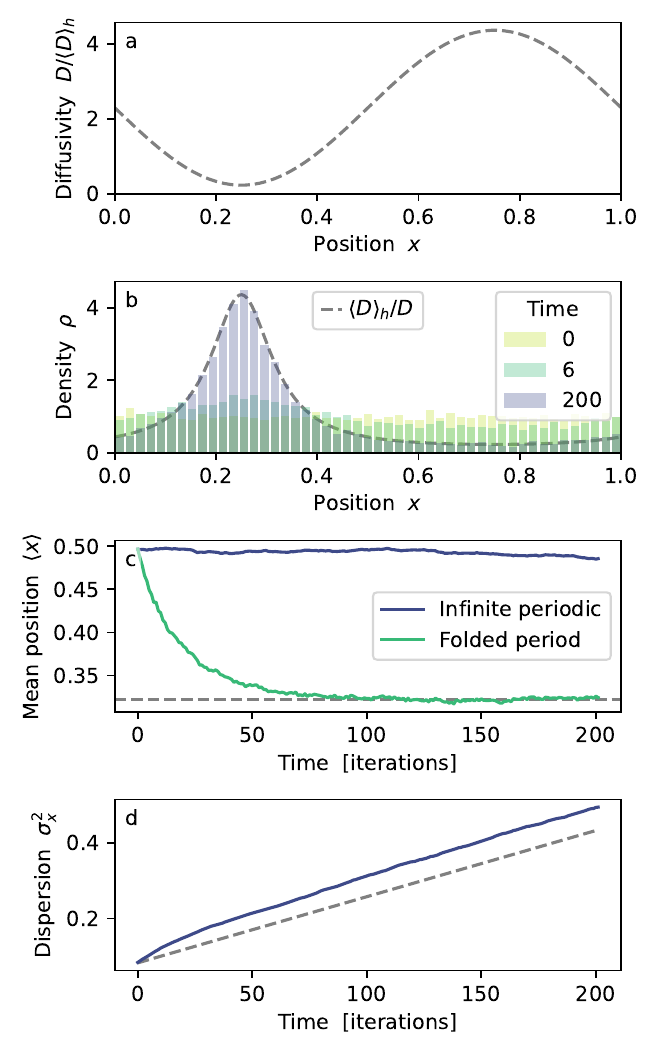}%
\caption{One-dimensional random walk simulation with heterogeneous diffusivity ($D \propto \ell^2$, equation~\eqref{eq:num_ell}, with $10^4$ walkers).
\figlet{a}: Diffusivity profile.
\figlet{b}: Position histogram at three different times, with folded periods (the walker's position is $x \mod{1}$). Dashed line shows theory, equation~\eqref{eq:one_over_D_normed}.
\figlet{c}: Evolution of the average position of the walkers. Dark blue line: Open, periodic system. Light green line: periods are folded over the interval $[0,1]$. Dashed line shows theoretical value, equation~\eqref{eq:conundrum}.
\figlet{d}: Dispersion of the walkers. Dashed line shows effective diffusion, equation~\eqref{eq:effective_diffusion}.
\label{fig:conundrum}}%
\end{figure}

To underline the peculiarity of equation~\eqref{eq:new_Fick}, we momentarily turn to numerical simulations. As it turns out, the simulation of a random walk that obeys equation~\eqref{eq:new_Fick} is simpler than one that yields Fick's law. One simply needs to use a random generator whose standard deviation depends on the current position of the walker, add its output to the walker's position, and repeat. Mathematically, one then simulates an explicit, unbiased random walk~\cite{volpe2016effective}---in mathematical terms, a martingale.

Figure~\ref{fig:conundrum} shows the results of such a simulation in one dimension. For illustration, we choose a step length that varies with the position $x$ of the particle  according to
\begin{equation}
  \ell = \sqrt{1 - 0.9 \sin (2\pi x)} \, ;
  \label{eq:num_ell}
\end{equation}
so that diffusivity, $D=\ell^2/(2\tau)$, depends sinusoidally on space (we assume that the time step is constant, figure~\ref{fig:conundrum}\figlet{a}). The simulation is made periodic, and thus $x$ is confined to the interval $[0,1]$. Starting with a uniform distribution of walkers, we find that they quickly gather around the diffusivity minimum, and their distribution then reaches a heterogeneous steady state (figure~\ref{fig:conundrum}\figlet{b}).

To put equation~\eqref{eq:one_over_D_normed} to the test, we simply compare the (normalized) equilibrium distribution of the walkers with $\langle D\rangle_h/D$, which is a function of $x$ (dashed line in figure~\ref{fig:conundrum}\figlet{b}). This expression matches the numerical distribution, thus showing that this simple simulation behaves like bouncing grains.

As the grains gather around the diffusivity minimum, their center of mass shifts. By definition, its position is $\langle x \rangle$, where $\langle \cdot \rangle$ is the ensemble average (formally, $\langle \cdot \rangle = \langle \rho \, \cdot  \rangle_a$ when the system is ergodic). Initially, the grains are uniformly distributed over the domain, that is, $\langle x \rangle=0.5$. As steady state is reached, this position shifts to
\begin{equation}
  \langle x \rangle = \dfrac{\int x / D \, \mathrm{d}x}{\int 1/D \, \mathrm{d}x} \approx  0.322 \, ,
  \label{eq:conundrum} 
\end{equation}
when steady state is reached, as visible in figure~\ref{fig:conundrum}\figlet{c}. This observation is just a direct consequence of equation~\eqref{eq:one_over_D_normed} but, when looked upon from a slightly different perspective, it can prove illuminating.

To see this, let us now release the walker from the domain $[0,1]$ and assume, instead, that it roams through an infinite, periodic diffusivity field, still defined by equation~\eqref{eq:num_ell}. One can think of a grain bouncing on an infinitely long, one-dimensional Chladni plate. As it turns out, the numerical simulation of this problem is exactly the same as before, only without folding the particle's position over a single period when we calculate its average position. We then find that the mean position $\langle x \rangle$ of the particle does not change (figure~\ref{fig:conundrum}\figlet{c}).

This apparent paradox is yet another way to distinguish equation~\eqref{eq:new_Fick} from Fick's law. Based on the former, indeed, the Smoluchowski equation reads
\begin{equation}
  \dfrac{\partial \rho}{\partial t} = \dfrac{\partial^2}{\partial x^2} \left(  D \rho \right) \, ,
  \label{eq:Smoluchowski}
\end{equation}
to which we need to adjoin boundary conditions. In the case of a walker in an infinite domain, the probability density and current vanish at infinity---provided the initial distribution is compact. Therefore, the average position of the particle reads
\begin{equation}
  \dfrac{\partial \langle x \rangle}{\partial t}
    = \int_{-\infty}^{\infty} x \dfrac{\partial^2}{\partial x^2} \left(  D \rho \right) \mathrm{d}x
    = - \left[D\rho\right]_{-\infty}^{\infty} = 0 \, ,
    \label{eq:average_x}
\end{equation}
were we have integrated by parts the first integral. This result accords with our simulation, but it contrasts with Fick's law, which induces a finite drift in a heterogeneous diffusivity field (appendix~\ref{sec:Fickian_drift}).

In the simulation presented here, the length of a step depends only on the walker's position at the onset of the jump. A direct consequence of this fact is that, at the onset of each step, the walker can go either left or right with the same probability, and with the same step length and velocity. Therefore, in the absence of any external force, its average position at step $i+1$ is the same as at step $i$---the definition of a martingale. Unless some boundary is reached, therefore, the average position of a grain cannot change (figure~\ref{fig:conundrum}\figlet{c}, dark blue line). In that sense, this process appears as the natural generalization of a random walk for heterogeneous diffusivity (we will mitigate this statement in section~\ref{sec:Markov}).

By contrast, when we introduce periodic boundaries in equation~\eqref{eq:average_x}, the integration by parts yields finite constants which allow the average position of the particle to drift (figure~\ref{fig:conundrum}\figlet{c}, light green line). This is because the initial distribution, mapped on an infinite periodic domain, now extends to infinity---it is not compact any more.

Returning to the infinite domain, we note that, although the average position of the walker does not change, it still disperses as diffusion goes on. Using again the Smoluchowski equation~\eqref{eq:Smoluchowski} to get the variance of the particle's position, we find
\begin{equation}
\begin{split}
  \frac{\partial \sigma_x^2}{\partial t}
    & = \int_{-\infty}^{\infty} x^2 \frac{\partial^2}{\partial x^2} \left(  D \rho \right) \mathrm{d}x \\
    & = 2\int_{-\infty}^{\infty} D \rho \, \mathrm{d}x
    = 2 \langle D \rangle \, .
  \end{split}
  \label{eq:variance_FP}
\end{equation}
Once the walkers have diffused over many periods of the diffusivity field, we may assume that their distribution is locally close to the equilibrium distribution~\eqref{eq:one_over_D}. Accordingly,
\begin{equation}
  \sigma_x^2 \sim 2 \langle D \rangle_h t
  \label{eq:effective_diffusion}
\end{equation}
and the effective diffusion coefficient is $\langle D \rangle_h$. This expression accords with numerical simulations (figure~\ref{fig:conundrum}\figlet{d}), and with the general treatment of this problem recently proposed by~\citet{giordano2024effective}. The fact that the effective diffusivity is the harmonic mean of diffusivity is yet another signature of the special kind of diffusion that equation~\eqref{eq:new_Fick} stands for.

This problem is reminiscent of a travel at varying velocity, the duration of which involves the harmonic mean of the velocity. Because the harmonic mean is always less than the arithmetic mean, the fastest diffusion of bouncing grains happens for uniform diffusivity, just like constant speed makes for the fastest travel. Conversely, introducing heterogeneity in the diffusivity creates traps which slow down the diffusion of the walker.

\subsection{Markov chain\label{sec:Markov}}

We now treat the bouncing of a grain explicitly as a discrete Markov process, wherein the length and duration of a jump depends only on the grain's starting position; in that sense, it is explicit. In one dimension, we may thus write, for $N$ steps,
\begin{equation}
  x_N = \sum_{i=1}^N \ell(x_{i}) \, w_i
  \label{eq:Ito}
\end{equation}
where $w_i$ is a random variable of vanishing mean and variance unity, which is independent of the grain's position. For illustration, we choose here $w_i=\pm 1$. Because the length of a jump depends only on the position $x_i$, and not on its next position $x_{i+1}$, this sum converges to an Itō integral in the limit of infinitesimal jumps. For this reason we hereafter refer to this process as ``Itō diffusion''.

By referring to an Itō process, we implicitly contrast it with the Stratonovitch integral, a discrete version of which would read
\begin{equation}
  x_N = \sum_{i=1}^N \dfrac{ \ell( x_{i} ) + \ell( x_{i+1} ) }{2} \, w_i  \, ,
  \label{eq:Stratonovitch}
\end{equation}
and with the anti-Itō integral---sometimes called ``isothermal’’ or named after \citet{hanggi1982nonlinear}---which reads 
\begin{equation}
  x_N = \sum_{i=1}^N \ell(x_{i+1}) \, w_i \, .
  \label{eq:anti-Ito}
\end{equation}
Both the Stratonovitch and anti-Itō integrals immediately appear unsuited to the random bouncing of a grain: How would the grain know on which substrate it will land at $x_{i+1}$ before hitting it \footnote{Stratonovitch and anti-Itō integrals are nonetheless the overdamped limit of different classes of dynamics \cite{van1992stochastic,volpe2016effective}.}? This question is of little relevance when the diffusivity is uniform, as the three sums---Itō, Stratonovitch and anti-Itō---converge to the same integral when the characteristics of the steps are uniform. However, when these characteristics depend on space, choosing equation~\eqref{eq:Ito}, \eqref{eq:Stratonovitch} or \eqref{eq:anti-Ito} changes the macroscopic diffusion process, even in the limit of a vanishing step length \cite{volpe2016effective}. The Itō integral yields equation~\eqref{eq:new_Fick}, whereas the anti-Itō one corresponds to Fick's law. As for the Stratonovich integral, it lies in between the two, and yields $\mathbf{j}=-\sqrt{D}\,\nabla(\sqrt{D}\rho)$. \citet[p.~231]{van1992stochastic} noted that this point, although well established mathematically, often remains a source of confusion.

We now return to equation~\eqref{eq:Ito}, which defines the displacement of the grain, and complement it with its counterpart in time, namely:
\begin{equation}
  t_N = \sum_{i=1}^N \tau(x_{i}) \, ,
  \label{eq:Ito_time}
\end{equation}
where $t_N$ is the duration of the trajectory. This expression, again, corresponds to an Itō process. An unusual fact about it is that, if the step duration $\tau$ is not constant, then the number of steps $N$ at a given time depends on the grain's trajectory---a consequence of equations~\eqref{eq:Ito} and \eqref{eq:Ito_time} being discretized in terms of individual bounces.

Let us first check that, on average, the grain does not move. By definition,
\begin{equation}
  \langle x_N \rangle = \sum_{i=1}^N \langle  \ell(x_{i}) \, w_i \rangle \, ,
  \label{eq:Ito_x_mean}
\end{equation}
and, since $w_i$ is uncorrelated with the grain's position,
\begin{equation}
  \langle  \ell(x_{i}) \, w_i \rangle = \langle  \ell(x_{i}) \rangle \, \langle w_i \rangle = 0 \, .
\end{equation}
Thus, for a fixed number of jumps $N$, the mean displacement $\langle x_N \rangle$ vanishes. It is not obvious, however, that this results holds for a fixed physical time $t$. Because $\tau$ varies from one jump to the next, indeed, the total number of jumps $N$ depends on the trajectory, according to equation~\eqref{eq:Ito_time}. The above derivation, however, can still serve as a heuristic explanation of the fact that an isolated Itō random walk is stationary on average, as straighforwardly established based on equation~\eqref{eq:new_Fick}.

As is customary for a random walk, we can also compute the variance of the grain's position:
\begin{equation}
  \langle x_N^2 \rangle
    = \left< \left( \sum_{i=1}^N \ell(x_{i}) \, w_i \right)^2 \right>
    = \left< \sum_{i=1}^N \ell(x_{i})^2 \right> \, ,
\end{equation}
where the last equality holds because all other terms are products of uncorrelated quantities. We now introduce the time step in the above sum:
\begin{equation}
  \langle x_N^2 \rangle
    = 2 \left< \sum_{i=1}^N  D(x_{i}) \tau(x_{i}) \right>
    = 2 \left<   \bar{ D } \, t_N \right> \, ,
    \label{eq:Ito_dispersion}
\end{equation}
where an overlined quantity is time-averaged over a trajectory. Again, this equation is exact for a fixed number of steps, but its validity when the physical time is fixed is unclear. In the continuous limit, however, one can expect it to become $ \langle x^2 \rangle = 2 \langle D \rangle t$, which is tantamount to equation~\eqref{eq:variance_FP} when ergodicity holds.

Because they introduce a correlation between the size of a step and its end point, neither the Stratonovitch sum~\eqref{eq:Stratonovitch} nor the anti-Itō one~\eqref{eq:anti-Ito} yield equations \eqref{eq:Ito_x_mean} and \eqref{eq:Ito_dispersion}---just like Fick's law did not yield equations~equation~\eqref{eq:average_x} and \eqref{eq:variance_FP} in section~\ref{sec:drift}. Instead, although this is less straightforward to show, only the anti-Itō formulation yields Fick's law, and therefore the moments associated to it \cite{wong1965convergence,van1992stochastic,volpe2016effective} (appendix~\ref{sec:Fickian_drift}). The above discussion thus supports the use of equation~\eqref{eq:new_Fick}, which corresponds to Itō diffusion, rather than Fick's law, to represent the diffusion of bouncing grains. The reason for this is that we can fairly represent the random walk of a bouncing grain as an explicit, unbiased process, wherein the noise manifests itself only when the grain hits the substrate, and vanishes while it is aloft.

The interpretation we propose thus departs from the conventional view that choosing the Itō, Stratonovitch or anti-Itō formalism is just a matter of mathematical convenience, since an additional drift term turns one into the other. Instead, following \citet[p.~236]{van1992stochastic} and \citet{volpe2016effective}, we prefer to interpret these formulations as the continuous limits of distinct microscopic models, that represent different physics. The drift term that distinguishes one from the other then takes on a physical meaning, and expresses itself visually in Chladni's experiment. From this perspective, bouncing grains are an instance of Itō diffusion, because the noise takes the form of singular events---an archetypal jump process.

This interpretation, hopefully, clarifies the debate between \citet{tailleur2008statistical}, who concluded that equation~\eqref{eq:new_Fick} could not describe a random walk, and \citet{schnitzer1993theory}, who treated the same equation as a possible representation of heterogeneous diffusion. The first authors, indeed, used a continuous mechanism to represent random walks, in which the probability to move or stop was a function of the \emph{current} particle position, thus excluding the possibility of Itō diffusion, wherein the particle carries the memory of its last jump over to the next bounce.

At this stage, it might be instructive to return to the conclusions of section~\ref{sec:drift}, and consider again the generality of Itō diffusion which, after the above, might appear as the unbiased---and therefore natural---generalization of homogeneous diffusion \footnote{This discussion owes much to A.P.~Petroff (personal communication).}. There is, however, no one-size-fits-all generalization. Brownian particles, for instance, are known to follow Fick's law, even when their diffusivity is heterogeneous \cite{faucheux1994confined}. If their motion is to be modeled with an overdamped random walk, therefore, it ought to be with the anti-Itō integral \cite{volpe2016effective}---unlike bouncing grains. In other words, Itō diffusion cannot represent inertial systems in contact with a thermal bath, even in the overdamped limit, because such systems must relax to the Boltzmann distribution.

\section{Heat flux\label{sec:heat}}

\begin{figure}[t]
\includegraphics[width=\figwidth]{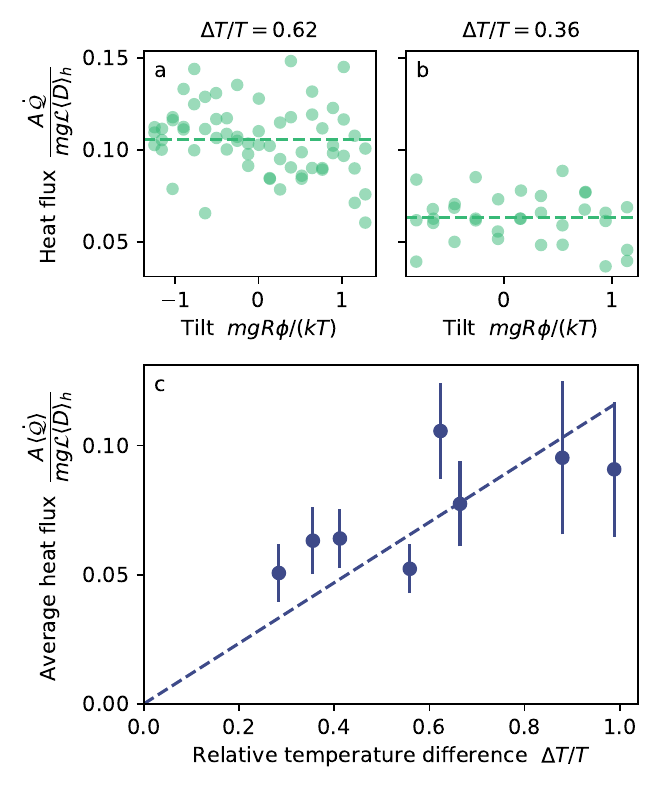}%
\caption{Heat flux from the soft substrate to the hard substrate.
\figlet{a}, \figlet{b}:~Dimensionless heat flux for experiments A and B, respectively. Each dot corresponds to a movie. Dashed line: average value.
\figlet{c}:~Average heat flux as a function of the relative temperature difference for experimental runs A to H. Dashed line: fitted linear relation with $\gamma=1.75$, equation~\eqref{eq:Q_dis}.
\label{fig:heat_flux}}%
\end{figure}

In the absence of any external force, bouncing grains gather in places of lesser diffusivity (section~\ref{sec:non-Fickian}), which are also colder places if equation~\eqref{eq:kT_D} holds. While a statistical equilibrium can then be reached ($\mathbf{j}=0$), it does not necessarily correspond to a thermodynamic equilibrium, wherein the heat flux $\mathbf{q}$ would vanish.

If, for illustration, we consider the heterogeneous substrate of figure~\ref{fig:substrate}\figlet{c}, we can say that, at statistical equilibrium and per unit time, the number of particles entering the area corresponding to the soft pad equals that of the particles leaving it (on average). Now, those that exit carry, on average, more kinetic energy than those that enter, thus causing a heat flux from the soft pad to the rigid disk. In accordance with intuition, thus, the heterogeneity of the substrate induces a heat flux from hot places to colder ones.

Of course, this heat flux is a key component of the thermodynamics of bouncing grains. In particular, the leakage of heat along the temperature gradient of the BL motor is a fundamental limitation on the latter's efficiency \cite{benjamin2008inertial}, because it cannot be separated from the driving mechanism. We now use our experiments to assess the heat flux carried by bouncing grains in statistical equilibrium, when no horizontal force acts on them.

% Assuming that each bouncing grain carries its kinetic energy, $kT$, over a jump, the average heat flux should read

We now introduce a phenomenological model for the transport of horizontal kinetic energy---which we call heat. We assume that, during a jump, a grain typically carries the kinetic energy $kT$ associated with its departure point. Combined with the particle flux law~\eqref{eq:new_Fick}, this suggests
\begin{equation}
  \mathbf{q} = - \nabla \left( D \rho  kT \right)
  \label{eq:q_j}
\end{equation}
where $\mathbf{q}$ is the heat flux parallel to the vibrating plate. This expression neglects possible correlations between the direction and the length of a jump, and the dissipation associated to flight and impact.

In statistical equilibrium, equation~\eqref{eq:one_over_D} implies that the heat flux becomes proportional to the temperature gradient:
\begin{equation}
  \mathbf{q} = - k \langle \rho \rangle \langle D \rangle_h \nabla T \, ,
  \label{eq:q_j_seq}
\end{equation}
where we have introduced $\langle \rho \rangle \equiv N/A$ to account for the total number of particles $N$, and the area $A$ of the domain they explore. We may then define the thermal conductivity as $k \langle \rho \rangle \langle D \rangle_h$, which is a constant for a given system in statistical equilibrium. The system thus uniformizes its thermal conductivity as it relaxes towards steady state.

The temperature gradient in equation~\eqref{eq:q_j_seq} becomes singular when the temperature field is discontinuous \cite{benjamin2008inertial}. This is a  problem in the present experiment, in which we assumed so far that the bouncing grains adjust their temperature instantly to local conditions. In reality, this adjustment most likely occurs over the course of a few bounces, that is, over a length of the order of the step size, which we write $\ell/\gamma$, thus defining the dimensionless relaxation parameter $\gamma$. In statistical equilibrium, the total exchange of heat $\dot{\cal Q}$ through the boundary of the soft pad then reads
\begin{equation}
  \dot{\cal Q} \approx  k \langle \rho \rangle \langle D \rangle_h \dfrac{ \gamma \Delta T }{ \ell } {\cal L}  \, ,
  \label{eq:Q}
\end{equation}
where $\cal L$ is the length of the interface between the two types of substrates, and $\Delta T$ is the (macroscopic) temperature difference between the two corresponding domains. Combining equations~\eqref{eq:tau}, \eqref{eq:ell} and \eqref{eq:v}, we can relate the step length to the temperature of a grain as
\begin{equation}
  \ell = \dfrac{4 kT \tan\theta}{mg} \, ,
\end{equation}
and thus rewrite the total heat flux \eqref{eq:Q} in a convenient form:
\begin{equation}
  \dot{\cal Q} \approx \langle \rho \rangle \langle D \rangle_h \gamma \dfrac{ m g {\cal L}  }{ 4 \tan \theta } \dfrac{\Delta T}{T} \, ,
  \label{eq:Q_dis}
\end{equation}
assuming that the relative change in temperature $\Delta T/T$ is not too large. 
Based on the grains' trajectories, we can measure every term in the above expression, and therefore estimate the value of the relaxation parameter $\gamma$.

For this, we need to evaluate the flux of kinetic energy through the boundary of the soft pad. We first collect all the segments in a trajectory that intersect the boundary of the soft pad, and record their direction $s_i$, as $s_i=1$ (or $s_i=-1$) if the grains enters (or leaves) the soft pad. We first check that the mass balance is statistically in equilibrium, that is
\begin{equation}
  \dfrac{1}{\sqrt{N}} \sum_{i=1}^N s_i \ll 1 
\end{equation}
where $N$ is the number of intersection for a single experimental movie. This ratio is $2\cdot 10^{-4}$ on average over all movies (and always less than $4\cdot 10^{-3}$), indicating that our data set is large enough to estimate equilibrium statistics.

For each intersection, we also record the length of the trajectory segment that crosses the boundary. Since the frame rate of the camera is fixed, this gives us a (horizontal) crossing velocity $v_i$, based on which we estimate the heat flux as
\begin{equation}
  \dot{\cal Q} = \dfrac{m}{2 \tau_m } \sum_{i=1}^N s_i v_i^2
\end{equation}
for each experimental movie of duration $\tau_m$. In figure~\ref{fig:heat_flux}\figlet{ab}, we represent this flux as a function of the tilt angle $\phi$, which was varied for runs A to H. We find that, despite a large scatter, we can estimate a statistically significant heat flux. Its value does not seem to be much affected by the tilt of the setup in our experiment, probably because the temperature gradient is orthogonal to the force induced by the tilt. We thus use all the data we have to get better statistics, thus assuming we can neglect the external force when we measure the heat flux. Plotting the mean value of this flux as a function of the relative temperature difference, we find that our observations are compatible with a linear relation (figure~\ref{fig:heat_flux}\figlet{c}). Fitting equation~\eqref{eq:Q_dis} to this data, we get an estimate of the relaxation parameter, $\gamma \approx 1.7\pm 0.5$. As expected, its value is of order one.

\section{Discussion}

In many respects, bouncing grains behave like Brownian particles. They diffuse like random walkers, and drift along external forces. Their temperature is well defined and they satisfy the fluctuation-dissipation relation. Their velocity also seems to follow the Maxwell-Boltzmann distribution. As long as their temperature is uniform, therefore, they can be treated as an analogue of classical thermodynamics---perhaps a useful one in the classroom.

Before bringing them it front of students, however, one should keep in mind that this system departs from classical statistical physics in subtle ways. As soon as the diffusivity of the grains varies, indeed, their macroscopic behavior reveals the nature of their microscopic dynamics---a series of repeated, dissipating collisions with the vibrating plate.

This was visualized even before the establishment of statistical physics by \citet{chladni1787entdeckungen}, whose experiment shows the departure from Boltzmann's statistics in the form of convoluted nodal lines. To phrase it metaphorically, in Chladni's experiment, the grains carry their inertial memory forward in time, and forget all about it at the next bounce. Due to this time asymmetry, they diffuse according to equation~\eqref{eq:new_Fick}, which allows for heterogeneous statistical equilibria that escape Fick's law. That the combination of path memory with its own erasure produces strange average dynamics should not surprise us---the bouncing droplets of \citet{couder2005walking} have been doing so for more than twenty years \cite{fort2010path}.

Due to the microscopic mechanism that draws them, Chladni figures are sustained at a cost. The vibrating substrate needs to supply the heat flux that necessarily arises from the heterogeneity of temperature. This flux follows a diffusion law that is consistent with equation~\eqref{eq:new_Fick}---at least, based on our observations. We interpret this disagreement with Fick's law as the macroscopic signature of an Itō random walk.

A remarkable property of the transport of heat by Itō diffusion is that, as the system reaches statistical equilibrium, its thermal conductivity becomes uniform, even though the density of grain does not, as shown by equation~\eqref{eq:q_j_seq}. This, again, would not be true with Fickian diffusion. Is this intriguing observation a coincidence, or does it reveal some deeper property of Itō diffusion? We would not venture any answer yet.

\bigskip

\begin{acknowledgments}
The idea of the paper arose during seminal discussions with S.~Protière. We are also grateful to E.~Lajeunesse and F.~Métivier for their help with the experiment, and to D.H.~Rothman, G.~Pucci, A.~Eddi, W.~Herreman, A.P.~Petroff, A.~Estevez-Torres, S.~Djambov, A.J.C.~Ladd and M.G.~Worster for inspiring discussions.

O.D was partially funded by PhysErosion ANR-22-CE30-0017.
\end{acknowledgments}

\appendix

\section{Acceleration of the plate\label{sec:acceleration}}

\begin{figure}[t]
\includegraphics[width=\figwidth]{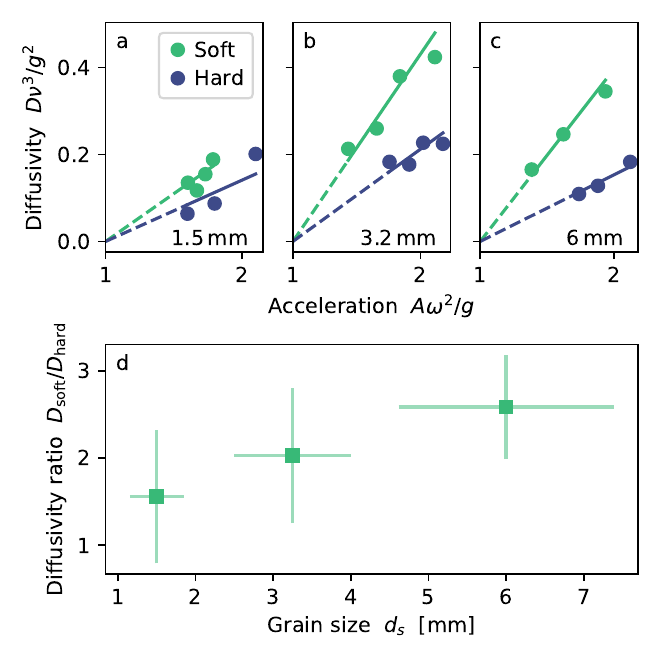}%
\caption{Dependence of the grains' diffusivity on the vibration intensity. \figlet{abc}: Increase of diffusivity with dimensionless acceleration, for runs I and N (\figlet{a}), J and K (\figlet{b}) and L and M (\figlet{c}). Straight lines show proportional relation fitted to the data. \figlet{d}: Dependence of the diffusivity ratio on grain size, based on the linear fit of \figlet{abc}.
\label{fig:soft_and_hard}}%
\end{figure}

Bouncing grains constantly dissipate energy, by friction with air and during their impact with the vibrating plate. The latter compensates for the energy loss by providing kinetic energy to the grains, in such a way as to maintain steady state. The mechanism by which this balance is reached is still unclear. It has attracted sustained attention in the simpler context of an elastic bead bouncing over a vibrating surface \cite{pieranski1983jumping,warr1996probability,chastaing2015dynamics}, due to its connection with the Feigenbaum cascade of bifurcations \cite{lichtenberg1980fermi} and, ultimately, with the Fermi–Pasta–Ulam–Tsingou problem of thermalisation \cite{ulam1961some}. Beyond a threshold acceleration ($A\omega^2/g$ large enough), such a bead bounces with an energy that increases with acceleration, perhaps linearly \cite{PhysRevE.68.031305}. This is qualitatively consistent with the behavior of bouncing grains, whose diffusivity also increases with acceleration, although the specific expression of this dependency is not yet known \cite{abramian2025chladni}.

We observe a similar behavior in the present experiment (figure~\ref{fig:soft_and_hard}\figlet{abc}). For any grain size, diffusivity increases with the acceleration of the plate, and it is always larger over the soft substrate than over the hard one. Our data set does not allow us to make this observation systematic, but we can plot the ratio of the soft diffusivity to the hard one, to find that it increases with grain size (figure~\ref{fig:soft_and_hard}\figlet{d}). We qualitatively interpret this observation as heavier grains being more sensitive to the softness of the elastomer---but we will not venture further than this speculation here. In this paper, we simply take its value as an experimental fact, which we use to interpret our observations.

\section{Fickian drift\label{sec:Fickian_drift}}

Here we calculate the drift of a Fickian particle in an heterogeneous diffusivity field, following the procedure of section~\ref{sec:drift}. We again assume that the particle density and current vanish at infinity. We thus find:
\begin{equation}
\begin{aligned}
  \dfrac{\partial \langle x \rangle}{\partial t}
    & = \int_{-\infty}^{\infty} x \, \dfrac{ \partial }{\partial x} \left( D \dfrac{\partial \rho}{\partial x} \right) \mathrm{d}x \\
    & = \int_{-\infty}^{\infty} \rho \, \dfrac{ \partial D }{\partial x} \, \mathrm{d}x
    = \left< \dfrac{ \partial D }{\partial x} \right>\\
\end{aligned}
\label{eq:average_x_Fick}
\end{equation}
after integrating by part twice. Unlike equation~\eqref{eq:new_Fick}, Fick's law induces a drift that is proportional to the gradient of diffusivity. On average, therefore, a Brownian particle moves toward places of higher diffusivity. This remarks shows that, when diffusivity is heterogeneous, Fick's law represents a biased random walk.

In the same fashion, we can now compute the spread of a parcel of Brownian particle in a heterogeneous diffusivity field, according to Fick's law:
\begin{equation}
\begin{aligned}
  \frac{\partial \sigma_x^2}{\partial t}
    & = \int_{-\infty}^{\infty} x^2 \, \dfrac{ \partial }{\partial x} \left( D \dfrac{\partial \rho}{\partial x} \right) \mathrm{d}x \\
    & = 2 \langle D \rangle + 2 \left< x \dfrac{ \partial D }{\partial x} \right> \, .\\
\end{aligned}
\label{eq:variance_FP_Fick}
\end{equation}
The variance on the particle's position thus increases linearly with time, as expected. This result is similar to equation~\eqref{eq:variance_FP}, but for an additional contribution proportional to the diffusivity gradient.

%%%%%%%%%%%%%%%%%%%%%%%%%%%%%%%%%%%%%%%%%%%%
\bibliography{./biblio_BL.bib}

@article{hanggi1982nonlinear,
  title={Nonlinear fluctuations: the problem of deterministic limit and reconstruction of stochastic dynamics},
  author={Hänggi, Peter},
  journal={Physical Review A},
  volume={25},
  number={2},
  pages={1130},
  year={1982},
  publisher={APS}
}

@article{volpe2016effective,
  title={Effective drifts in dynamical systems with multiplicative noise: a review of recent progress},
  author={Volpe, Giovanni and Wehr, Jan},
  journal={Reports on Progress in Physics},
  volume={79},
  number={5},
  pages={053901},
  year={2016},
  publisher={IOP Publishing}
}

@article{raja2024diffusive,
  title={Diffusive lensing as a mechanism of intracellular transport and compartmentalization},
  author={Raja Venkatesh, Achuthan and Le, Kathy H and Weld, David M and Brandman, Onn},
  journal={Elife},
  volume={12},
  pages={RP89794},
  year={2024},
  publisher={eLife Sciences Publications, Ltd}
}

@article{giordano2024effective,
  title={Effective diffusion constant of stochastic processes with spatially periodic noise},
  author={Giordano, Stefano and Blossey, Ralf},
  journal={Physical Review E},
  volume={110},
  number={4},
  pages={044123},
  year={2024},
  publisher={APS}
}

@article{fort2010path,
  title={Path-memory induced quantization of classical orbits},
  author={Fort, Emmanuel and Eddi, Antonin and Boudaoud, Arezki and Moukhtar, Julien and Couder, Yves},
  journal={Proceedings of the National Academy of Sciences},
  volume={107},
  number={41},
  pages={17515--17520},
  year={2010},
  publisher={National Academy of Sciences}
}

@article{couder2005walking,
  title={Walking and orbiting droplets},
  author={Couder, Yves and Protiere, Suzie and Fort, Emmanuel and Boudaoud, Arezki},
  journal={Nature},
  volume={437},
  number={7056},
  pages={208--208},
  year={2005},
  publisher={Nature Publishing Group UK London}
}

@inproceedings{ulam1961some,
  title={On some statistical properties of dynamical systems},
  author={Ulam, SM},
  booktitle={Proc. 4th Berkeley Sympos. Math. Statist. and Prob},
  volume={3},
  pages={315--320},
  year={1961}
}

@article{lichtenberg1980fermi,
  title={Fermi acceleration revisited},
  author={Lichtenberg, AJ and Lieberman, MA and Cohen, RH},
  journal={Physica D: Nonlinear Phenomena},
  volume={1},
  number={3},
  pages={291--305},
  year={1980},
  publisher={Elsevier}
}

@article{pieranski1983jumping,
  title={Jumping particle model. Period doubling cascade in an experimental system},
  author={Piera{\'n}ski, P},
  journal={Journal de Physique},
  volume={44},
  number={5},
  pages={573--578},
  year={1983},
  publisher={Soci{\'e}t{\'e} fran{\c{c}}aise de physique}
}

@article{warr1996probability,
  title={Probability distribution functions for a single-particle vibrating in one dimension: Experimental study and theoretical analysis},
  author={Warr, S and Cooke, W and Ball, RC and Huntley, JM},
  journal={Physica A: Statistical Mechanics and its Applications},
  volume={231},
  number={4},
  pages={551--574},
  year={1996},
  publisher={Elsevier}
}

@incollection{risken1989fokker,
  title={Fokker-planck equation},
  author={Risken, Hannes},
  booktitle={The Fokker-Planck equation: methods of solution and applications},
  pages={63--95},
  year={1989},
  publisher={Springer}
}

@article{PhysRevE.68.031305,
  title = {Energy of a single bead bouncing on a vibrating plate: Experiments and numerical simulations},
  author = {G\'eminard, J.-C. and Laroche, C.},
  journal = {Phys. Rev. E},
  volume = {68},
  issue = {3},
  pages = {031305},
  numpages = {5},
  year = {2003},
  month = {Sep},
  publisher = {American Physical Society},
  doi = {10.1103/PhysRevE.68.031305},
  url = {https://link.aps.org/doi/10.1103/PhysRevE.68.031305}
}

@article{chastaing2015dynamics,
  title={Dynamics of a bouncing ball},
  author={Chastaing, J-Y and Bertin, Eric and G{\'e}minard, J-C},
  journal={American Journal of Physics},
  volume={83},
  number={6},
  pages={518--524},
  year={2015},
  publisher={AIP Publishing}
}

@misc{munkres,
  title = {{Munkres (Hungarian) algorithm for the assignment problem in Python}},
  howpublished = {\url{https://pypi.org/project/munkres/}},
}

@misc{browntrack,
  title = {{BrownTrack Python library}},
  howpublished = {\url{https://github.com/odevauchelle/BrownTrack/tree/main}},
}

@misc{gphoto,
  title = {{gPhoto$^2$ digital-camera Unix software}},
  howpublished = {\url{http://gphoto.org/}},
}

@article{scikit-image,
 title = {scikit-image: image processing in {P}ython},
 author = {van der Walt, {S}t\'efan and {S}ch\"onberger, {J}ohannes {L}. and
           {Nunez-Iglesias}, {J}uan and {B}oulogne, {F}ran\c{c}ois and {W}arner,
           {J}oshua {D}. and {Y}ager, {N}eil and {G}ouillart, {E}mmanuelle and
           {Y}u, {T}ony and the scikit-image contributors},
 year = {2014},
 month = {6},
 volume = {2},
 pages = {e453},
 journal = {PeerJ},
 issn = {2167-8359},
 url = {https://doi.org/10.7717/peerj.453},
 doi = {10.7717/peerj.453},
 note = {\url{https://scikit-image.org/}}
}

@article{faucheux1994confined,
  title={Confined brownian motion},
  author={Faucheux, Luc P and Libchaber, Albert J},
  journal={Physical Review E},
  volume={49},
  number={6},
  pages={5158},
  year={1994},
  publisher={APS}
}

@article{schnitzer1993theory,
  title={Theory of continuum random walks and application to chemotaxis},
  author={Schnitzer, Mark J},
  journal={Physical Review E},
  volume={48},
  number={4},
  pages={2553},
  year={1993},
  publisher={APS}
}

@article{tailleur2008statistical,
  title={Statistical mechanics of interacting run-and-tumble bacteria},
  author={Tailleur, Julien and Cates, Michael E},
  journal={Physical review letters},
  volume={100},
  number={21},
  pages={218103},
  year={2008},
  publisher={APS}
}

@article{berger2009optimal,
  title={Optimal potentials for temperature ratchets},
  author={Berger, Florian and Schmiedl, Tim and Seifert, Udo},
  journal={Physical Review E},
  volume={79},
  number={3},
  pages={031118},
  year={2009},
  publisher={APS}
}

@article{benjamin2008inertial,
  title={Inertial effects in B{\"u}ttiker-Landauer motor and refrigerator at the overdamped limit},
  author={Benjamin, R and Kawai, R},
  journal={Physical Review E},
  volume={77},
  number={5},
  pages={051132},
  year={2008},
  publisher={APS}
}

@article{companionPRL,
  title={{A granular B{\"u}ttiker-Landauer motor}},
  author={Devauchelle, O. and Popović, P. and Szymczak, P. and Abramian, A. and Lazarus, A.},
  journal={Physical Review Letters},
  year={2025},
  publisher={APS}
}

@article{abramian2025chladni,
  title={Chladni patterns explained by the space-dependent diffusion of bouncing grains},
  author={Abramian, Ana{\"\i}s and Proti{\`e}re, Suzie and Lazarus, Arnaud and Devauchelle, Olivier},
  journal={Physical Review Research},
  volume={7},
  number={3},
  pages={L032001},
  year={2025},
  publisher={APS}
}

@article{wong1965convergence,
  title={On the convergence of ordinary integrals to stochastic integrals},
  author={Wong, Eugene and Zakai, Moshe},
  journal={The Annals of Mathematical Statistics},
  volume={36},
  number={5},
  pages={1560--1564},
  year={1965},
  publisher={JSTOR}
}

@book{van1992stochastic,
  title={Stochastic processes in physics and chemistry},
  author={Van Kampen, Nicolaas Godfried},
  volume={1},
  year={1992},
  publisher={Elsevier}
}

@article{van1988relative,
  title={Relative stability in nonuniform temperature},
  author={Van Kampen, NG},
  journal={IBM Journal of Research and Development},
  volume={32},
  number={1},
  pages={107--111},
  year={1988},
  publisher={IBM}
}

@article{landauer1988motion,
  title={Motion out of noisy states},
  author={Landauer, Rolf},
  journal={Journal of statistical physics},
  volume={53},
  number={1},
  pages={233--248},
  year={1988},
  publisher={Springer}
}

@article{faraday1831xvii,
  title={On a peculiar class of acoustical figures; and on certain forms assumed by groups of particles upon vibrating elastic surfaces},
  author={Faraday, M.},
  journal={Philosophical transactions of the Royal Society of London},
  number={121},
  pages={299--340},
  year={1831},
  publisher={The Royal Society London}
}

@article{kudrolli2008swarming,
  title={Swarming and swirling in self-propelled polar granular rods},
  author={Kudrolli, Arshad and Lumay, Geoffroy and Volfson, Dmitri and Tsimring, Lev S},
  journal={Physical review letters},
  volume={100},
  number={5},
  pages={058001},
  year={2008},
  publisher={APS}
}

@article{arango2016stochastic,
  title={{Stochastic models for Chladni figures}},
  author={Arango, Jaime and Reyes, Carlos},
  journal={Proceedings of the Edinburgh Mathematical Society},
  volume={59},
  number={2},
  pages={287--300},
  year={2016},
  publisher={Cambridge University Press}
}

@article{buttiker1987transport,
  title={Transport as a consequence of state-dependent diffusion},
  author={B{\"u}ttiker, M},
  journal={Zeitschrift f{\"u}r Physik B Condensed Matter},
  volume={68},
  number={2},
  pages={161--167},
  year={1987},
  publisher={Springer}
}

@article{grabec2017vibration,
  title={{Vibration driven random walk in a Chladni experiment}},
  author={Grabec, Igor},
  journal={Physics Letters A},
  volume={381},
  number={2},
  pages={59--64},
  year={2017},
  publisher={Elsevier}
}

@book{chladni1787entdeckungen,
  title={{Entdeckungen {\"u}ber die Theorie des Klanges}},
  author={Chladni, Ernst Florens Friedrich},
  year={1787},
  address={Leipzig},
  publisher={Bey Weidmanns Erben Und Reich},
}
%%%%%%%%%%%%%%%%%%%%%%%%%%%%%%%%%%%%%%%%%%%%

%%%%%%%%%%%%%%%%%%%%%%%%%%%%%%%%%%%%%%%%%%%%
\end{document}